\documentclass[11pt]{cernrep}
\usepackage{graphicx}

\newcommand{\lesssim}{\raise.3ex\hbox{$<$\kern-.75em\lower1ex\hbox{$\sim$}}}
\newcommand{\grtssim}{\raise.3ex\hbox{$>$\kern-.75em\lower1ex\hbox{$\sim$}}}

\begin{document}

\title{COLOR SUPERCONDUCTIVITY IN DENSE QUARK MATTER}

\author{Igor A. Shovkovy\thanks{~~On leave from
                 Bogolyubov Institute for Theoretical Physics,
                 03143, Kiev, Ukraine}}

\institute{Frankfurt Institute for Advanced Studies, 
Johann Wolfgang Goethe-Universit\"at\\
D-60438 Frankfurt am Main, Germany}

\maketitle

\begin{abstract}
A brief introduction into the properties of dense quark matter 
is given. Recently proposed gapless color superconducting phases 
of neutral and beta-equilibrated dense quark matter are discussed. 
The current status in the field is described, and the promising 
directions of the future research are outlined.
\end{abstract}

\section{Introduction}
\label{Intro}

At sufficiently high baryon density, matter is expected to be 
deconfined. The physical degrees of freedom in a deconfined 
phase are quarks and gluons, rather than usual hadrons. At 
present the theory cannot predict reliably where in the QCD 
phase diagram the corresponding deconfinement transition should 
occur. The issue gets further complicated by the fact that the
deconfinement is not associated with a symmetry-related order 
parameter and, thus, does not need to be marked by any real phase 
transition. Leaving aside this well-known conceptual difficulty, 
here I discuss the recent progress in the studies of cold and 
dense matter which, owing in part to the property of the 
asymptotic freedom in QCD, allows a relatively rigorous
treatment. 

It was suggested long time ago that quark matter may exist inside the
central regions of compact stars \cite{quark-star}. By making use of
the property of asymptotic freedom in QCD \cite{GWP}, it was argued
that quarks interact weakly, and that realistic calculations taking
full account of strong interactions are possible for sufficiently
dense matter \cite{ColPer}. The argument of Ref.~\cite{ColPer}
consisted of the two main points: (i) the long-range QCD interactions
are screened in dense medium causing no infrared problems, and
(ii) at short distances, the interaction is weak enough to allow
the use of the perturbation theory. As will become clear below, 
the real situation in dense quark matter is slightly more subtle.

\section{Color superconductivity}
\label{CSC}

By assuming that very dense matter is made of weakly
interacting quarks, one could try to understand the thermodynamic
properties of the corresponding ground state by first completely
neglecting the interaction between quarks. In order to construct
the ground state, it is important to keep in mind that quarks are 
fermions, i.e., particles with a half-integer spin, $s=1/2$. They
should obey the Pauli exclusion principle which prohibits for two 
identical fermions to occupy the same quantum state.

In the ground state of non-interacting quark matter at zero 
temperature, quarks occupy all available quantum states with the
lowest possible energies. This is formally described by the 
following quark distribution function:
\begin{equation}
f_F(\mathbf{k}) =\theta\left(\mu-E_\mathbf{k}\right), 
\quad \mbox{at} \quad T=0,
\label{f_F_T=0}
\end{equation}
where $\mu$ is the quark chemical potential, and $E_\mathbf{k}\equiv
\sqrt{k^2+m^2}$ is the energy of a free quark (with mass $m$) in the
quantum state with the momentum $\mathbf{k}$ (by definition, $k\equiv
|\mathbf{k}|$). As one can see, $f_F(\mathbf{k})=1$ for the states
with $k<k_F\equiv \sqrt{\mu^2-m^2}$, indicating that all states with
the momenta less than the Fermi momentum $k_F$ are occupied. The
states with the momenta greater than the Fermi momentum $k_F$ are
empty, i.e., $f_F(\mathbf{k})=0$ for $k>k_F$.

It appears that the perturbative ground state of quark matter,
characterized by the distribution function in Eq.~(\ref{f_F_T=0}),
is unstable when there is an attractive (even arbitrarily weak in
magnitude!) interaction between quarks. This is because of the
famous Cooper instability \cite{Cooper}. The instability develops 
as a result of the formation of Cooper pairs $\langle q_{\mathbf{k}}\,
q_{-\mathbf{k}}\rangle$ made of quarks from around the highly
degenerate Fermi surface, i.e., quarks with the absolute value 
of momenta $k\simeq k_F$. Such Cooper pairs are bosonic states, 
and they occupy the same lowest energy quantum state at $T=0$, 
producing a version of a Bose condensate. The corresponding 
ground state of quark matter is then a superconductor. This 
is similar to the ground state of an electron system in the 
Bardeen-Cooper-Schrieffer (BCS) theory of low-temperature
superconductivity \cite{BCS}. Of course, some qualitative 
differences also arise because quarks, unlike electrons, 
come in various flavors (e.g., up, down and strange) and 
carry non-Abelian color charges. To emphasize the difference, 
superconductivity in quark matter is called {\it color 
superconductivity}. For recent review on color superconductivity 
see Ref.~\cite{reviews}.

As in low-temperature superconductors in solid state physics, 
one of the main consequences of color superconductivity in 
dense quark matter is the appearance of a nonzero gap in the 
one-particle energy spectrum. In the simplest case, the 
dispersion relation of gapped quasiparticles is given by
\begin{equation}
E_{\Delta}(k)= \sqrt{\left(E_{\mathbf{k}}-\mu\right)^2+\Delta^2},
\label{E_qp}
\end{equation}
where $\Delta$ is the gap. The presence of a nonzero gap affects 
kinetic (e.g., conductivities and viscosities) as well as 
thermodynamic (e.g., the specific heat and the equation of 
state) properties of quark matter \cite{reviews}. 

Historically, it has been known for a rather long time that dense 
quark matter should be a color superconductor \cite{BarFra,Bail}. 
In many studies in the past this fact was commonly ignored, however. 
Only recently, the potential importance of this phenomenon was 
appreciated. To large extent, this has been triggered by the observation 
\cite{cs} that the value of the color superconducting gap $\Delta$ can
be as large as $100\,\mbox{MeV}$ at baryon densities existing in the 
central regions of compact stars, i.e., at densities which are a 
few times larger than the normal nuclear density, $n_0 \simeq 0.15 
\mbox{~fm}^{-3}$. A posteriori, of course, this estimate is hardly 
surprising within the framework of QCD, in which the energy scale is 
set by $\Lambda_{\rm QCD} \simeq 200\,\mbox{MeV}$. Yet this observation 
was very important because the presence of a large energy gap 
in the quasiparticle spectrum may allow to extract signatures of 
color superconducting matter in observational data from compact 
stars.

\section{Two-flavor color superconductivity ($N_f=2$)}

The simplest color superconducting phase is the two-flavor color
superconductor (2SC). This is a color superconducting phase in 
quark matter made of up and down quarks. 

In weakly interacting regime of QCD at asymptotic densities, the 2SC
phase of matter was studied from first principles in Ref.~\cite{weak}.
It should be mentioned, however, even at the highest densities 
existing in the central regions of compact stars ($n\lesssim 10n_0$) 
quark matter is unlikely to be truly weakly interacting. In such a 
situation, the use of the microscopic theory of strong interactions 
is very limited, and one has to rely on various effective models of 
QCD. A very simple type of such a model, used for the description of 
color superconducting matter, is the Nambu-Jona-Lasinio (NJL) model 
with a local four-fermion interaction (for a review see, e.g., 
Ref.~\cite{Buballa}). One of its simplest versions is defined by the 
following Lagrangian density \cite{SKP}:
\begin{eqnarray}
{\cal L}_{\rm NJL} &=& 
\bar\psi_i^a\left(i\gamma^\mu \partial_\mu
+\gamma^0\mu -m^{(0)}_i\right) \psi_i^a
+ G_S\left[(\bar\psi \psi)^2
+ (i\bar\psi \gamma_5\vec{\tau}\psi)^2\right]
\nonumber\\ 
&+&G_D (i \bar{\psi}^C \varepsilon \epsilon^a \gamma_5 \psi )
 (i \bar{\psi} \varepsilon \epsilon^a \gamma_5 \psi^C),
\label{NJL-action}
\end{eqnarray}
where $\psi^C=C \bar{\psi}^T$ is the charge-conjugate spinor and
$C=i\gamma^2\gamma^0$ is the charge conjugation matrix. The matrix
$C$ is defined so that $C\gamma_{\mu} C^{-1}=-\gamma_{\mu}^{T}$.
Regarding the other notation, $\vec{\tau}=(\tau^1,\tau^2,\tau^3)$ are
the Pauli matrices in the flavor space, while $(\varepsilon)^{ik} \equiv
\varepsilon^{ik}$ and $(\epsilon^a)^{bc} \equiv \epsilon^{abc}$ are the
antisymmetric tensors in the flavor and in the color spaces, respectively.
The dimensionful coupling constant $G_S=5.01\,\mbox{GeV}^{-2}$ and the
momentum integration cutoff parameter $\Lambda=0.65\,\mbox{GeV}$ (which
appears only in loop calculations) are adjusted so that the values
of the pion decay constant and the value of the chiral condensate
take their standard values in vacuum QCD: $F_\pi = 93$ MeV and
$\langle\bar{u}u\rangle=\langle\bar{d}d\rangle =(-250\,\mbox{MeV})^3$
\cite{SKP}. Without loss of generality, the strength of the coupling 
constant $G_D$ is taken to be proportional to the value of $G_S$: 
$G_D=\eta G_S$ where $\eta$ is a dimensionless parameter of order 1. 
It is important that $\eta$ is positive, which corresponds to 
an attraction in the color-antisymmetric diquark channel. This 
property is suggested by the microscopic interaction in QCD at 
high density, as well as by the instanton-induced interaction 
at low density \cite{cs}.

The color-flavor structure of the condensate of spin-0 Cooper 
pairs in the 2SC phase reads
\begin{equation}
\left\langle \left(\bar{\psi}^C\right)_i^a \gamma^5 \psi_j^b
\right\rangle
\sim \varepsilon_{ij} \epsilon^{abc} .
\label{2sc-cond}
\end{equation}
In a fixed gauge, the color orientation of this condensate can 
be chosen arbitrarily. It is conventional to point the condensate 
in the third (blue) color direction, $\left\langle 
\left(\bar{\psi}^C\right)_i^a \gamma^5 \psi_j^b \right\rangle 
\sim \varepsilon_{ij} \epsilon^{ab3}$. Then, the Cooper pairs 
in the 2SC phase are made of the red and green quarks only,
while blue quarks do not participate in pairing at all. These
unpaired blue quarks give rise to ungapped quasiparticles in 
the low-energy spectrum of the theory.

The flavor antisymmetric structure in Eq.~(\ref{2sc-cond}) corresponds
to a singlet representation of the global SU(2)$_L\times$SU(2)$_R$ chiral 
group. This means that the (approximate) chiral symmetry is not broken in
the 2SC ground state. In fact, there are no other global continuous
symmetries which are broken in the 2SC phase. There exist, however,
several approximate symmetries which are broken. One of them is the
approximate U(1)$_A$ symmetry which is a good symmetry at high density
when the instantons are screened \cite{scr_inst}. Its breaking in the
2SC phase results in a pseudo-Nambu-Goldstone boson \cite{low-e-2sc}.
Additional four pseudo-Nambu-Goldstone states may appear as a
result of a less obvious approximate axial color symmetry 
discussed in Ref.~\cite{pseudoNG}.

In the ground state, the vector-like SU(3)$_c$ color gauge group
is broken down to the SU(2)$_c$ subgroup. Therefore, five out
of total eight gluons of SU(3)$_c$ become massive due to the
Anderson-Higgs mechanism. The other three gluons, which correspond
to the unbroken SU(2)$_c$, do not interact with the gapless blue
quasiparticles. They give rise to low-energy SU(2)$_c$ gluodynamics.
The red and green quasiparticles decouple from this low-energy
SU(2)$_c$ gluodynamics because they are gapped \cite{RSS}.

The gap equation in the NJL model in the mean field approximation
looks as follows:
\begin{equation}
\Delta \simeq \frac{4 G_D}{\pi^2} \int_0^{\Lambda}
\left(\frac{\Delta}{\sqrt{(p-\mu)^2+\Delta^2}}
+\frac{\Delta}{\sqrt{(p+\mu)^2+\Delta^2}}\right)
p^2 d p .
\label{gap-eq-NJL}
\end{equation}
This gap equation is analogous to the Schwinger-Dyson equation
in QCD \cite{weak} in which the gluon long-range interaction is 
replaced by a local interaction.

The approximate solution to the gap equation in 
Eq.~(\ref{gap-eq-NJL}) reads
\begin{equation}
\Delta \simeq 2\sqrt{\Lambda^2-\mu^2}
\exp\left(-\frac{\pi^2}{8G_D\mu^2}+\frac{\Lambda^2-3\mu^2}{2\mu^2}\right).
\end{equation}
This is very similar to the BCS solution in the case of low 
temperature superconductivity in solid state physics \cite{BCS}. 
As in the BCS theory, it has the same type non-analytic dependence 
on the coupling constant and the same type dependence on the density 
of quasiparticle states at the Fermi surface. (Note that in QCD 
at asymptotic density, in contrast, the long-range interaction 
leads to a qualitatively different non-analytic dependence of the 
gap on the coupling constant, $\Delta\sim \mu \, \alpha_s^{-5/2}\exp 
\left(-C/\sqrt{\alpha_s}\right)$ where $C=3(\pi/2)^{3/2}$ 
\cite{weak}).

When the quark chemical potential $\mu$ takes a value in the range
between $400\,\mbox{MeV}$ and $500\,\mbox{MeV}$, and the strength
of the diquark pairing is $G_D=\eta G_S$ with $\eta$ between $0.7$ and
$1$, the value of the gap appears to be of order $100\,\mbox{MeV}$.
In essence, this is the result that was obtained in Ref.~\cite{cs}.

\section{Color-flavor locked phase ($N_f=3$)}

It may happen that dense baryonic matter is made not only of
the lightest up and down quarks, but of strange quarks as well.
In fact, because of a possible reduction in the free energy
from converting non-strange quarks into strange quarks, one may
even speculate that strange quark matter is the true ground state of
baryonic matter \cite{Bod}.

The constituent strange quark mass in vacuum QCD is estimated to be
of order $500\,\mbox{MeV}$. Its current mass is about $100\,\mbox{MeV}$.
In dense baryonic matter in stars, therefore, the strange quark mass 
should be somewhere in the range between the two limits, $100\,\mbox{MeV}$ 
and $500\,\mbox{MeV}$. It is possible then that strange quarks also 
participate in Cooper pairing.

Let me first discuss an idealized version of three-flavor quark matter,
in which all quarks are assumed to be massless. A more realistic
case of a nonzero strange quark mass will be discussed briefly
in Secs.~\ref{regimes} and \ref{g2CSgCFL}. In the massless case, the quark model
possesses the global SU(3)$_L\times$SU(3)$_R$ chiral symmetry and the
global U(1)$_B$ symmetry connected with the baryon number conservation.
This is in addition to SU(3)$_c$ color gauge symmetry. Note that the
generator $Q=\mbox{diag}_{\rm flavor}(\frac{2}{3},-\frac{1}{3},
-\frac{1}{3})$ of the U(1)$_{\rm em}$ symmetry of electromagnetism
is traceless, and therefore it coincides with one of the vector-like
generators of the SU(3)$_L\times$SU(3)$_R$ chiral group.

To large extent, the color and flavor structure of the spin-0 diquark
condensate of Cooper pairs in the three-flavor quark matter is
fixed by the symmetry of the attractive diquark channel and the Pauli 
exclusion principle. In particular, this is given by the following 
ground state expectation value \cite{cfl}:
\begin{equation}
\left\langle \left(\bar{\psi}^C\right)_i^a \gamma^5 \psi_j^b
\right\rangle
\sim \sum_{I,J=1}^{3} c^{I}_{J}\varepsilon_{ijI} \epsilon^{abJ}
+\cdots,
\label{cfl-cond}
\end{equation}
which is antisymmetric in the color and flavor indices of the
constituent quarks, cf. Eq.~(\ref{2sc-cond}). The $3\times 3$
matrix $c^{I}_{J}$ is determined by the global minimum of the free
energy. It appears that $c^{I}_{J}=\delta^{I}_{J}$. The ellipsis
on the right hand side stand for a contribution which is symmetric
in color and flavor. A small contribution of this type is always
induced in the ground state, despite the fact that it corresponds
to a repulsive diquark channel \cite{cfl,weak-cfl}. This is not
surprising after noting that the symmetric condensate, i.e., 
$\left\langle\left(\bar{\psi}^C\right)_i^a \gamma^5 \psi_j^b \right\rangle
\sim \delta_{i}^{a}\delta_{j}^{b}+\delta_{j}^{a}\delta_{i}^{b}$,
does not break any additional symmetries \cite{cfl}.

In the ground state, determined by the condensate (\ref{cfl-cond}),
the chiral symmetry is broken down to its vector-like subgroup.
The mechanism of this symmetry breaking is rather unusual, however.
To see this clearly, it is helpful to rewrite the condensate as 
follows:
\begin{equation}
\left\langle \psi_{L,i}^{a,\alpha} \epsilon_{\alpha\beta}
\psi_{L,j}^{b,\beta}\right\rangle
=-\left\langle \psi_{R,i}^{a,\dot\alpha} \epsilon_{\dot\alpha\dot\beta}
\psi_{R,j}^{b,\dot\beta}\right\rangle
\sim \sum_{I=1}^{3} \varepsilon_{ijI} \epsilon^{abI}
+\cdots,
\label{LL-RR-cond}
\end{equation}
where $\alpha,\beta,\dot\alpha,\dot\beta=1,2$ are the spinor indices.
The condensate of left-handed fields, $\langle \psi_{L,i}^{a,\alpha} 
\epsilon_{\alpha\beta}\psi_{L,j}^{b,\beta}\rangle$, breaks
the SU(3)$_c$ color symmetry and the SU(3)$_L$ chiral symmetry, but
leaves the diagonal SU(3)$_{L+c}$ subgroup unbroken. Indeed, as one
can check, this condensate remains invariant under a flavor 
transformation ($g_L$) and a properly chosen compensating color 
transformation ($g_c=g_L^{-1}$). Similarly, the condensate of 
right-handed fields,
$\langle \psi_{R,i}^{a,\dot\alpha} \epsilon_{\dot\alpha\dot\beta}
\psi_{R,j}^{b,\dot\beta}\rangle$, leaves the SU(3)$_{R+c}$ subgroup 
unbroken.

When both condensates are present, the symmetry of the ground 
state is given by the diagonal subgroup SU(3)$_{L+R+c}$. This is 
because one has no freedom to use two different compensating color 
transformations. At the level of global symmetries, the original 
SU(3)$_{L}\times$SU(3)$_{R}$ symmetry of the model is broken
down to the vector-like SU(3)$_{L+R}$, just like in vacuum. (Note, 
however, that the CFL phase is superfluid because the global $U(1)_{B}$ 
symmetry is broken by the diquark condensate in the ground state.)
Unlike in vacuum, the chiral symmetry breaking does {\em not} result from 
any condensate mixing left- and right-handed fields. Instead, it results
primarily from two separate condensates, made of left-handed fields  
and of right-handed fields only. The flavor orientations of the two
condensates are ``locked'' to each other by color transformations.
This mechanism is called locking, and the corresponding phase of
matter is called color-flavor-locked (CFL) phase \cite{cfl}.

The gap equation in the three-flavor quark matter is qualitatively 
the same as in the two-flavor case. The differences come only
from a slightly more complicated color-flavor structure of the
off-diagonal part of the inverse quark propagator (gap matrix)
\cite{cfl,weak-cfl},
\begin{equation}
\Delta^{ij}_{ab}=i\gamma^5\left[
\frac{1}{3}\left(\Delta_1+\Delta_2\right)\delta^{i}_{a}\delta^{j}_{b}
-\Delta_2\delta^{i}_{b}\delta^{j}_{a}\right],
\label{delta_cfl}
\end{equation}
where two parameters $\Delta_1$ and $\Delta_2$ determine the values
of the gaps in the quasiparticles spectra. In the ground state, which
is invariant under the SU(3)$_{L+R+c}$ symmetry, the original nine quark
states give rise to a singlet and an octet of quasiparticles with
different values of the gaps in their spectra. When a small 
color-symmetric diquark condensate is neglected, one finds
that the gap of the singlet ($\Delta_1$) is twice as large as 
the gap of the octet ($\Delta_2$), i.e., $\Delta_1=2\Delta_2$. In 
general, however, this relation is only approximate. In QCD 
at asymptotic density, the dependence of the gaps on the 
quark chemical potential was calculated in 
Refs.~\cite{weak-cfl,loops}.

\section{Dense matter inside stars}
\label{matter-in-stars}

As discussed in Secs.~\ref{Intro} and \ref{CSC}, it is natural to expect 
that color superconducting phases may exist in the interior of compact
stars. The estimated central densities of such stars might be sufficiently
large for producing deconfined quark matter. Then, such matter develops
the Cooper instability and becomes a color superconductor. It should
also be noted that typical temperatures inside compact stars are 
so low that a spin-0 diquark condensate, if produced, would not melt.
(Of course, this may not apply to a short period of the stellar 
evolution immediately after the supernova explosion.)

In the preceding sections, only idealized versions of dense matter, in 
which the Fermi momenta of pairing quarks were equal, were discussed.
These cannot be directly applied to a realistic situation that is thought  
to occur inside compact stars. The reason is that matter in the bulk of a 
compact star should be neutral (at least, on average) with respect to 
electric as well as color charges. Also, matter should remain in $\beta$ 
(chemical) equilibrium, i.e., the $\beta$ processes $d \to u + e^{-} + 
\bar\nu_{e}$ and $u + e^{-} \to d + \nu_{e}$ (as well as $s \to u + e^{-} 
+ \bar\nu_{e}$ and $u + e^{-} \to s + \nu_{e}$ in the presence of strange 
quarks) should go with equal rates. (Here it is assumed that there is no 
neutrino trapping in stellar matter. In the presence of neutrino trapping,
the situation changes \cite{gCFL-nu}. Also, the situation changes in the 
presence of a very strong magnetic field \cite{magCFL}, but the discussion 
of its effect is outside the scope of this short review.)

Formally, $\beta$ equilibrium is enforced by introducing a set of chemical 
potentials ($\mu_{i}$) in the partition function of quark matter,
\begin{equation}
Z=\mbox{Tr}\,\exp\left(-\frac{H+\sum_i \mu_{i} Q_i}{T}\right).
\end{equation}
The total number of independent chemical potentials $\mu_{i}$ is equal 
to the number of conserved charges $Q_i$ in the model. For example, in 
two-flavor quark matter, it suffices to consider only three relevant 
conserved charges: the baryon number $n_B$, the electric charge $n_Q$, 
and the color charge $n_{8}$. (Note that these may not be sufficient in a 
general case \cite{note}.) Then, the matrix of quark chemical potentials 
is given in terms of the baryon chemical potential (by definition, 
$\mu_B\equiv 3\mu$), the electron chemical potential ($\mu_e$) and 
the color chemical potential ($\mu_8$) \cite{no2sc,g2SC,other},
\begin{equation}
\hat\mu_{ij, \alpha\beta}= (\mu \delta_{ij}- \mu_e Q_{ij})
\delta_{\alpha\beta} + \frac{2}{\sqrt{3}}\mu_{8} \delta_{ij}
(T_{8})_{\alpha \beta},
\end{equation}
where $Q$ and $T_8$ are the generators of U(1)$_{\rm em}$ of
electromagnetism and the U(1)$_{8}$ subgroup of the gauge group 
SU(3)$_{c}$.

The other important condition in stellar matter is that of 
charge neutrality. In order to get an impression regarding the 
importance of charge neutrality in a large macroscopic chunk 
of matter, such as a core of a compact star, one can estimate 
the corresponding Coulomb energy. A simple calculation leads 
to the following result: 
\begin{equation}
E_{\rm Coulomb} \sim n_Q^2 R^5 \sim 10^{26} M_\odot c^2
\left(\frac{n_Q}{10^{-2} e/\mbox{fm}^3}\right)^2
\left(\frac{R}{1\,\mbox{km}}\right)^5,
\label{E_Coulomb}
\end{equation}
here $R$ is the radius of the quark matter core, whose charge density
is denoted by $n_Q$. It is easy to see that this energy is not an
extensive quantity: the value of the corresponding {\em energy density}
increases with the size of the system as $R^2$. By taking a typical
value of the charge density in the 2SC phase, $n_Q\sim 10^{-2}
e/\mbox{fm}^3$, the energy in Eq.~(\ref{E_Coulomb}) becomes a factor
of $10^{26}$ larger than the rest mass energy of the Sun! To avoid
such an incredibly large energy price, the charge neutrality $n_Q=0$
should be satisfied with a very high precision.

In the case of two-flavor quark matter, one can argue that the neutrality
is achieved if the number density of down quarks is approximately
twice as large as number density of up quarks, $n_d \approx 2n_u$.
This follows from the fact that the negative charge of the down quark
($Q_d=-1/3$) is twice as small as the positive charge of the up quark
($Q_u=2/3$). When $n_d \approx 2n_u$, the total electric charge
density is vanishing in absence of electrons, $n_Q\approx Q_d n_d
+Q_u n_u\approx 0$. It turns out that even a nonzero density of electrons,
required by the $\beta$ equilibrium condition, could not change this
relation much.

The argument goes as follows. One considers noninteracting massless
quarks. In $\beta$ equilibrium, the chemical potentials of the up quark
and the down quark, $\mu_u$ and $\mu_d$, should satisfy the relation
$\mu_d=\mu_u+\mu_e$ where $\mu_e$ is the chemical potential of
electrons (i.e., up to a sign, the chemical potential of the
electric charge). By assuming that $\mu_d\approx 2^{1/3}\mu_u$, i.e.,
$n_d\approx2n_u$ as required by the neutrality in absence of electrons,
one obtains the following result for the electron chemical potential:
$\mu_e=\mu_d-\mu_u \approx 0.26\mu_u$. The corresponding
density of electrons is $n_e\approx 6\cdot 10^{-3}n_u$, i.e., 
$n_e\ll n_u$ which is in agreement with the original assumption 
that $n_d\approx 2n_u$ in neutral matter.

While the approximate relation $n_d\approx 2n_u$ may be slightly
modified in an interacting system, the main conclusion remains
qualitatively the same. The Fermi momenta of up and down quarks, whose
pairing is responsible for color superconductivity, are generally
non-equal when neutrality and $\beta$ equilibrium are imposed. 
This affects the dynamics of Cooper pairing and, as a consequence,
some color superconducting phases may become less favored than 
others. For example, it is argued in Ref.~\cite{no2sc}, that a 
mixture of unpaired 
strange quarks and the non-strange 2SC phase, made of up and 
down quarks, is less favorable than the CFL phase after the
charge neutrality condition is enforced. In addition, it was 
found that neutrality and $\beta$ equilibrium may give rise 
to new unconventional pairing patterns \cite{g2SC,gCFL}.

\section{Different dynamical regimes in neutral matter}
\label{regimes}

By studying neutral two-flavor quark matter, it was found that
there exist three qualitatively different dynamical regimes, defined 
by the (largely unknown) strength of diquark coupling \cite{g2SC}.
Similar regimes were also suggested to exist in three-flavor 
quark matter when the strange quark mass is not negligibly small 
\cite{gCFL,breach}. (Other effects due to a non-zero strange 
quark mass are discussed in Ref.~\cite{cfl+mesons}.)

The simplest regime corresponds to weak diquark coupling. In 
this case, cross-flavor Cooper pairing of quarks with non-equal
Fermi momenta is energetically unfavorable. The ground state of 
neutral matter corresponds to the normal phase. This would be 
precisely the case in QCD at asymptotic density if there existed 
only up and down quark flavors. (Formally, this is also the case 
when there are six quark flavors as in the Standard Model!) One 
should note, however, that a much weaker spin-1 pairing between 
quarks of same flavor is not forbidden in such neutral matter. 
In fact, spin-1 condensates would be inevitable if the temperature 
is sufficiently low.

The other limiting case is the strongly coupled regime. It is clear
that, if the value of the diquark coupling is sufficiently large, the
color condensation could be made as strong as needed to overcome a
finite mismatch between the Fermi surfaces of pairing quarks. In 
this regime, the ground state is in the 2SC/CFL phase because 
$\beta$-equilibrium and charge neutrality have little effect. 

It turns out that there also exists an intermediate regime, in which
the diquark coupling is neither too weak nor too strong. It was proposed 
that the ground state in this regime is given by the so-called gapless 
superconductor \cite{g2SC,gCFL}, briefly discussed in the next 
section.

\section{Gapless 2SC and CFL phases}
\label{g2CSgCFL}

Without going into details, the characteristic feature of a gapless 
superconducting phase is the existence of gapless quasiparticle 
excitations in its low-energy spectrum. The simplest examples are 
the gapless 2SC (g2SC) \cite{g2SC} and gapless CFL (gCFL) 
\cite{gCFL} phases. In the g2SC case, for example, there exists 
a doublet of quasiparticles with the following dispersion 
relation \cite{g2SC}:
\begin{equation}
E_{\Delta}(k)=\sqrt{\left(E_{\mathbf{k}}-\bar{\mu}\right)^2+\Delta^2}
-\delta\mu,
\end{equation}
where $\Delta$ is the value of the gap parameter, $\bar{\mu}\equiv 
(\mu_1+\mu_2)/2$ is the average chemical potential and $\delta\mu
\equiv (\mu_1-\mu_2)/2$ is the mismatch between the chemical 
potentials of pairing quarks. When $\Delta<\delta\mu$, it takes 
vanishingly small amount of energy to excite quasiparticles with
momenta in the vicinity of $k_{\pm}\equiv \bar{\mu}\pm 
\sqrt{(\delta\mu)^2-\Delta^2}$. Similar quasiparticles also 
exist in gCFL phase as well.

When the g2SC and gCFL phases were suggested, it was argued that 
their thermodynamic stability was enforced by the charge neutrality 
condition \cite{g2SC}. In a homogeneous macroscopic system, such a 
condition is necessary in order to avoid a huge energy price due to 
the Coulomb long-range interaction. Remarkably, this condition has
no analogue in solid state physics. Thus, one argued that the known 
problems of the so-called Sarma \cite{Sarma} phase may not apply to 
the g2SC/gCFL phases.

\section{Chromomagnetic instability and suggested alternatives}

Rather quickly, it was discovered that the gapless phases have 
problems of their own \cite{pi}. Namely, the screening Meissner 
masses of several gauge bosons are imaginary in the ground state, 
indicating a new type (chromomagnetic) instability in quark 
matter. The original calculation was performed for the g2SC 
phase \cite{pi}, but a similar observation regarding the 
gCFL phase was also made soon \cite{pi-gcfl,pi-gcfl-KF}.

In the case of the g2SC phase, e.g., it was found that the screening 
Meissner masses for five out of total eight bosons are imaginary 
when $0<\Delta/\delta\mu<1$. In addition and most surprisingly, 
it was also found that four gauge bosons have imaginary masses
even in the {\em gapped} 2SC phase when $1<\Delta/\delta\mu<\sqrt{2}$. 
The most natural interpretation of these results is that the 
instability might be resolved through the formation of a gluon 
condensate in the ground state \cite{pi}. It is fair to note, 
however, that the exact nature of the instability (and in the
case of $1<\Delta/\delta\mu<\sqrt{2}$, in particular) is still 
poorly understood. The presence of the imaginary masses even 
in the {\em gapped} phase (i.e., when $1<\Delta/\delta\mu<\sqrt{2}$), 
may suggest that the gapless superconductivity is not the only 
reason for the instability. While there remain many open questions, 
a partial progress in resolving the problem has already been made 
\cite{LOFF-inst,gluonic}.

In the gCFL phase, the instability is seen only for three gauge bosons
\cite{pi-gcfl-KF}. The corresponding screening Meissner masses have a 
dependence on the mismatch parameter which is similar to the that for 
the 8th gluon in the g2SC phase. The fate of such an instability
has not been clarified completely. At asymptotic density, however,
it was suggested that the stable ground state might be given by a 
phase with an additional p-wave meson condensate \cite{p-wave}. 
Whether a similar phase also exists in two-flavor quark matter 
is unclear because the situation is further complicated there by 
(i) the absence of a natural mesonic state among the low-energy 
excitations and (ii) the onset of the ``abnormal'' chromomagnetic 
instability for the gluons $A^{4}_{\mu},\, A^{5}_{\mu},\, A^{6}_{\mu}$,
and $A^{7}_{\mu}$. Instead of a p-wave meson condensate, the so-called 
``gluonic'' phase may be realized \cite{gluonic}.

The presence of the chromomagnetic instability in g2SC and gCFL 
phases indicates that these phases cannot be stable ground states 
of matter. It should be emphasized, however, that this does not 
mean that, in nature, gapless phases are ruled out completely. First 
of all, there is an indication from studies in non-relativistic models
that similar instabilities may not appear under some special 
conditions \cite{breach-stab,ss}. In addition, most of the alternatives
to the g2SC \cite{g2SC} and gCFL \cite{gCFL} phases, that have been 
suggested \cite{LOFF-inst,gluonic,p-wave}, share the same qualitative 
feature: their spectra of low-energy quasiparticles possess gapless 
modes. In fact, this seems to be not accidental but the most natural 
outcome of a very simple observation: the ordinary ``gapped'' versions 
of superconductivity are hardly consistent with the unconventional 
Cooper pairing, required in neutral and $\beta$-equilibrated quark 
matter.

\section{Discussion}

In conclusion, 
there has been a tremendous progress in recent studies of dense 
baryonic matter. This started from a seemingly innocuous 
observation that the size of the gap in the energy spectrum 
of color superconducting quark matter, under conditions realized 
in stars, could be of the same order as the QCD scale \cite{cs}. 
This opened a whole new chapter in studies of new states of dense 
matter that could exist inside compact stars. In addition to a 
phenomenological/observational interest, the recent studies in 
color superconductivity in neutral and $\beta$-equilibrated 
matter revealed a wide range of fundamentally new possibilities 
stemming from unconventional Cooper pairing. It is plausible that 
in the future a cross-disciplinary importance of this finding may 
even overshadow its role in physics of compact stars.

If color superconducting quark matter indeed exists in the 
interior of compact stars, it should affect some important 
transport and thermodynamic properties of stellar matter 
which may, in turn, affect some observational data from 
stars. Among the most promising signals are the cooling rates 
\cite{cooling-sp0,cooling-sp1} and the rotational slowing down 
of stars \cite{rotation}. Also, new states of matter could affect 
the stellar mass-radius relation \cite{EoS}, and even lead to the 
existence of a new family of compact stars \cite{new-class}. Color 
superconductivity can also affect directly as well as indirectly
many other observed properties of stars. In some cases, for example,
superconductivity may be accompanied by baryon superfluidity and/or 
the electromagnetic Meissner effect. If matter is superfluid,
rotational vortices would be formed in the stellar core, and they
would carry a portion of the angular momentum of the star. Because
of the Meissner effect, the star interior could become threaded
with magnetic flux tubes. In either case, the star evolution may
be affected. While some studies on possible effects of color 
superconductivity in stars have already been attempted, the 
systematic study remains to be done in the future.

The development in the field also resulted in obtaining reliable 
nonperturbative solutions in QCD at asymptotic densities 
\cite{weak,weak-cfl,loops,cfl+mesons}, shedding some light on the 
structure of the QCD phase diagram in the regime inaccessible by
lattice calculations. By itself, this has a fundamental theoretical 
importance. Also, this may provide valuable insights into 
the theory of strong interactions. One of the examples might 
be the idea of duality between the hadronic and quark description 
of QCD \cite{cont}. In the future, the structure of the QCD 
phase diagram and the properties of various color superconducting 
phases should be studied in more detail. While many different 
phases of quark matter have been proposed, there is no certainty 
that all possibilities have already been exhausted.

\section*{Acknowledgments}
I would like to thank the organizers of ``Extreme QCD'' in 
Swansea for organizing an interesting workshop, and for 
creating a warm and stimulating atmosphere. This work was 
supported in part by the Virtual Institute of the Helmholtz 
Association under grant No. VH-VI-041 and by Gesellschaft 
f\"{u}r Schwerionenforschung (GSI) and by Bundesministerium 
f\"{u}r Bildung und Forschung (BMBF).

\end{document}